\documentclass[12pt]{iopart}
\usepackage{graphicx}
\usepackage{cite}
\usepackage{amsfonts}
\usepackage{wrapfig}

\begin{document}

\title[Entropy-based Tuning of Musical Instruments]
      {Entropy-based Tuning of Musical Instruments}
\author{Haye Hinrichsen}
\address{Universit\"at W\"urzburg\\
	 Fakult\"at f\"ur Physik und Astronomie\\
         D-97074 W\"urzburg, Germany}
\ead{hinrichsen@physik.uni-wuerzburg.de}

\begin{abstract}
The human sense of hearing perceives a combination of sounds `in tune' if the corresponding harmonic spectra are correlated, meaning that the neuronal excitation
pattern in the inner ear exhibits some kind of order. Based on this observation it is suggested that musical instruments such as pianos can be tuned by minimizing
the Shannon entropy of suitably preprocessed Fourier spectra. This method reproduces not only the correct stretch curve but also similar pitch fluctuations as in
the case of high-quality aural tuning. 
\end{abstract}

\parskip 2mm 

\section{Introduction}

\begin{figure}
\centering\includegraphics[width=130mm]{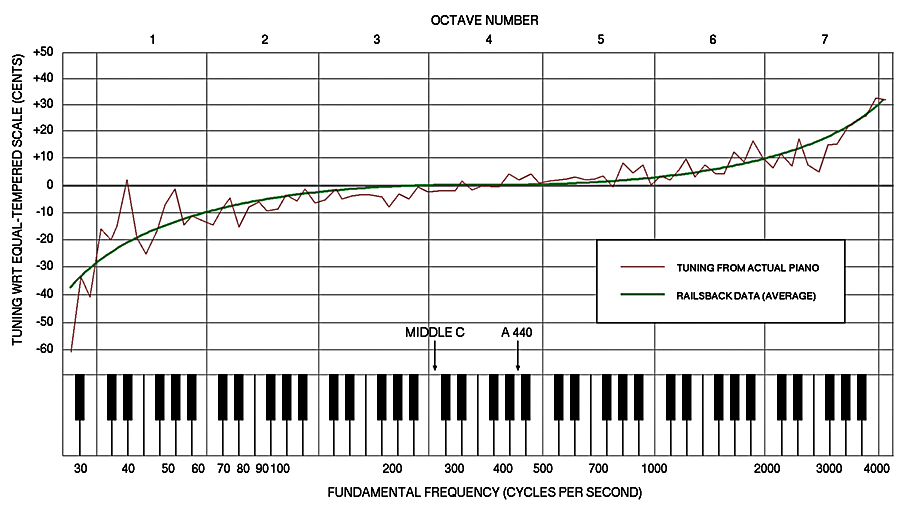}
\caption{Typical tuning curve of a piano (figure taken from~\cite{figuresource}). The plot shows how much the fundamental frequency of each note deviates
from the equal-tempered scale. The deviations are measured in \textit{cents}, defined as 1/100 of a halftone which corresponds to a frequency ratio of $2^{1/1200} \approx 1.0005778$.}
\label{figrailsback}
\end{figure}

Western musical scales are based on the \textit{equal temperament} (ET), a system of tuning in which adjacent notes differ
by a constant frequency ratio of $2^{1/12}$~\cite{roederer}. Tuning musical instruments in equal temperament by ear used to be a challenging task which
was carried out by iterating cyclically over certain interval sequences. Today this task is performed much more accurately with the help of electronic
tuning devices which automatically recognize the tone, measure its frequency, and display the actual pitch deviation from the theoretical value.
However, if one uses such a device to tune a piano or a harpsichord exactly in equal temperament, the instrument as a whole will eventually sound as if it were
out of tune, even though each string is tuned to the correct frequency. This surprising effect was first explained by O.~L.~Railsback in 1938,
who showed that this perception is caused by inharmonic corrections in the overtone spectrum~\cite{railsback}.
Professional aural tuners compensate this inharmonicity by small deviations, a technique known as \textit{stretching}. The stretch depends on the specific amount
of inharmonicity and can be visualized in a tuning chart (see Fig.~\ref{figrailsback}).

Since the inharmonicity changes from instrument to instrument, it is difficult to compute the appropriate stretch by means of electronic tuning methods.
Some appliances allow the user to preselect typical average stretches for certain classes and sizes of instruments. More advanced tuning devices measure
the individual overtone spectra of all notes and compute the necessary stretch by correlating higher harmonics. Although the latter method gives fairly
good results and is increasingly used by professional piano tuners, many musicians are still convinced that electronic tuning cannot compete with high-quality
aural tuning by a skilled piano technician. This raises the question why aural tuning is superior to electronic methods.

When measuring the frequencies of an aurally well-tuned piano one finds that the tuning curve is not smooth: rather it exhibits irregular fluctuations from note to note
on top of the overall stretch (see Fig.~\ref{figrailsback}). At first glance one might expect that these fluctuations are randomly distributed and caused by the natural
inaccuracy of human hearing. However, as we will argue in the present paper, these fluctuations are probably not totally random; they might instead reflect to some extent
the individual irregularities in the overtone spectra of the respective instrument and thus could play an essential role in high-quality tuning. Apparently our ear can
find a better compromise in the highly complex space of spectral lines than most electronic tuning devices can do.

As a first step towards a better understanding of these fluctuations, we suggest that a musical instrument can be tuned by minimizing an appropriate entropy functional.
This hypothesis anticipates that a complex sound is perceived as `pleasant', `harmonic' or `in tune' if the corresponding neuronal activity is ordered in such a way that
the Shannon entropy of the excitation pattern is minimal. The hope is that such an entropy-based optimization allows one to find a better compromise between slightly
detuned harmonics than a direct comparison of selected spectral lines. 

\newpage

\section{Harmonic spectrum, musical scales, and temperaments}

\begin{wrapfigure}{r}{70mm}
\vspace{-8mm}
\begin{center}
\includegraphics[width=60mm]{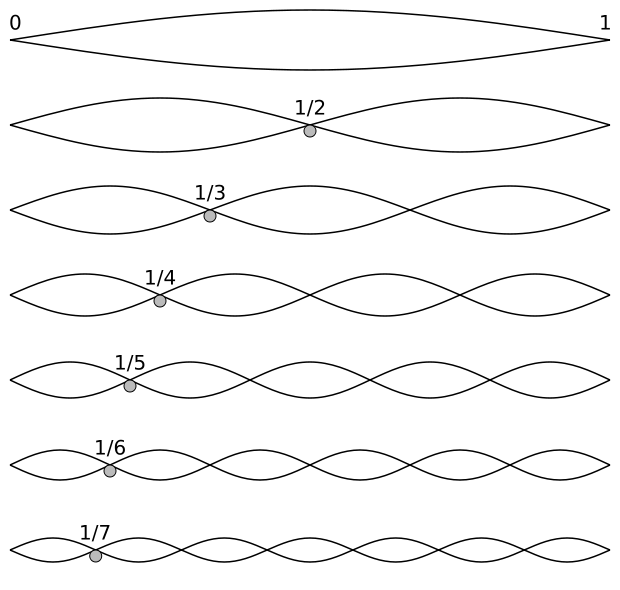} 
\caption{}Harmonic modes of a string~\cite{figuresource2}.
\end{center}
\vspace{-10mm}
\end{wrapfigure}

Sound waves produced by musical instruments involve many Fourier components. The simp\-lest example is the spectrum of a vibrating string~\cite{FletcherRossing}.
Depending on the excitation mechanism, one finds not only the fundamental mode with the frequency $f_1$ but also a large number of higher partials (overtones) with frequencies
$f_2,f_3,\ldots,$. For an ideal string the frequencies of the higher partials are just multiples of the fundamental frequency, i.e.
\begin{equation}
f_n \;=\; n f_1\,. 
\end{equation}
Such a linearly organized spectrum of overtones is called \textit{harmonic}. 

\begin{figure}
\includegraphics[width=150mm]{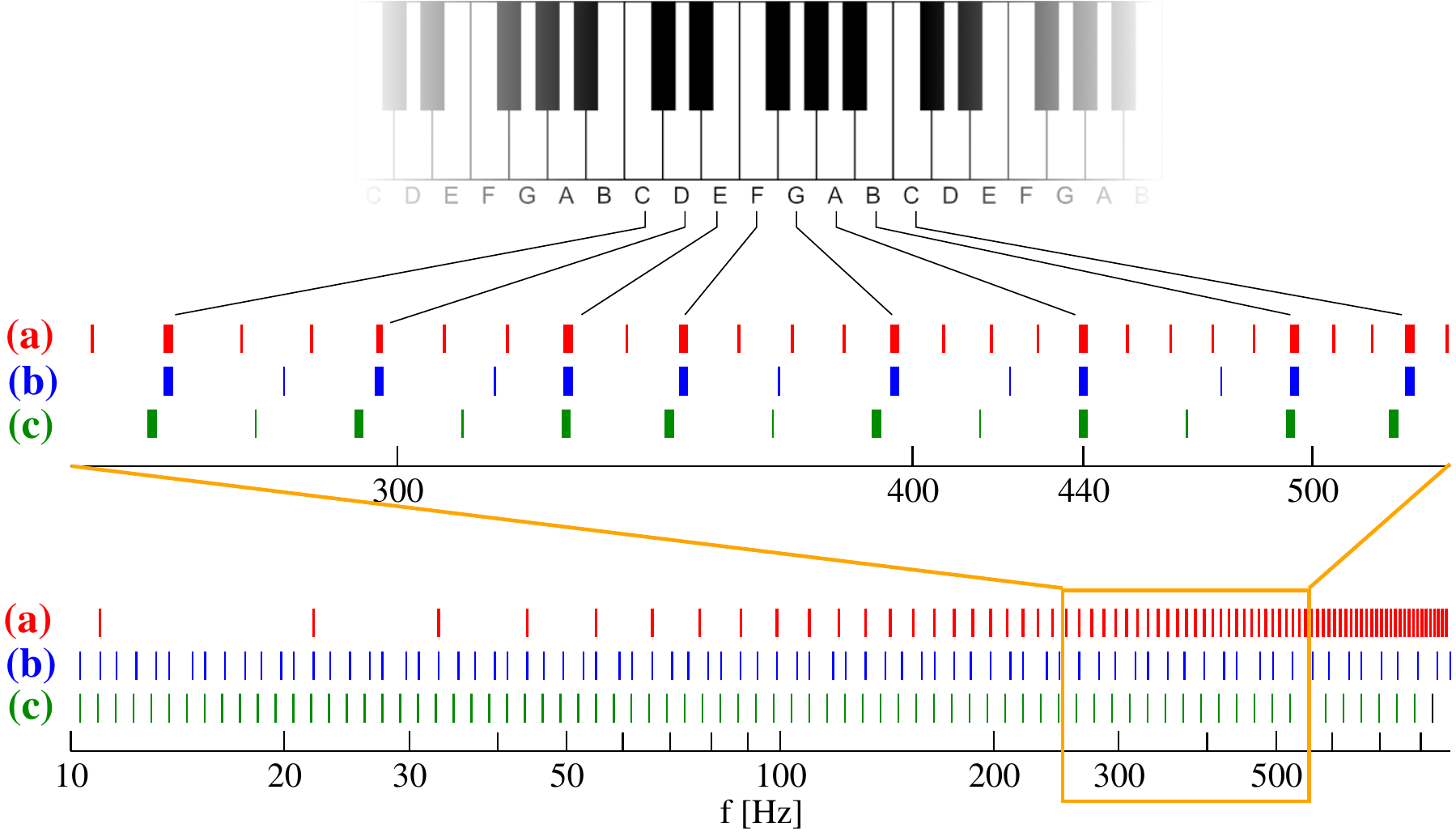}
\caption{\small Harmonic overtone spectrum compared with musical temperaments (logarithmic scale). Lower part: (a) Fundamental frequency $f_1=11$ Hz and the corresponding series of
harmonics. (b) Just intonation in C-Major with octave-repeating patterns of non-equidistant pitches. (c) Equally tempered intonation with constant pitch
differences. The upper part of the figure shows a zoom of the octave C4-C5. As can be seen, the heptatonic scale (the white piano keys, thick lines) of the just
intonation (b) matches perfectly with the harmonics in (a), while the half tones (black keys, thin lines) do not lock in. Contrarily the equal temperament (c) deviates
in all tones except A440 but it is equidistant and therefore invariant under shifts (transpositions) of the musical key. }
\label{figscale}
\end{figure}

Since harmonic overtones are ubiquitous in Nature, our sense of hearing prefers intervals with simple frequency ratios for which the spectra of overtones partially coincide. Examples are the octave (2:1), the perfect fifth (3:2), and the perfect fourth (4:3), which play an important role in any kind of music. On the other hand, music is usually based on scales of notes arranged in \textit{octave-repeating} patterns. Since the frequency doubles from octave to octave, it grows exponentially with the index of the notes. This exponentially organized structure of octave-repeating notes is in immediate conflict with the linear spectrum of the harmonics. A \textit{musical scale} can be seen as the attempt to reconcile these conflicting schemes, defining the frequencies in such a way that the harmonics of a given note coincide as much as possible with other notes of the scale in higher octaves. As demonstrated in Fig.~\ref{figscale}, this leads quite naturally to \textit{heptatonic scales} with seven tones per octave, on which most musical cultures are based. In traditional Western music the seven tones (the white piano keys) are supplemented by five halftones (black keys), dividing an octave in twelve approximately equal intervals.

As the twelve intervals establish a compromise between the arithmetically ordered harmonics and the exponentially organized musical scale, their sizes are not uniquely
given but may vary in some range. Over the centuries this freedom has led to the development of various tuning schemes, known as \textit{intonations} or \textit{temperaments}, which approximate the harmonic series to a different extent. One extreme is the \textit{just intonation}, which is entirely built on simple rational frequency ratios. As shown in Fig.~\ref{figscale}b, the just intonation shares many spectral lines with a suitable harmonic spectrum. However, this tuning scheme is not equidistant on a logarithmic representation, breaking translational invariance under key shifts (transpositions). Therefore, the just intonation is in tune only with respect to a specific musical key (e.g. C Major), while it is out of tune in all other musical keys.

With the increasing complexity of Western music and the development of advanced keyboard instruments such as harpsichords, organs and pianos,
more flexible intonations were needed, where the musical key can be changed without renewed tuning. Searching for a better compromise between purity
(rational frequency ratios) and temperament independence (transposition invariance) various tuning schemes have been developed, including the famous meantone temperament
in the renaissance and the well-tempered intonation of the baroque era. Since the 19$^{\rm th}$ century Western music is predominantly based on the aforementioned
equal temperament, which is fully invariant under key shifts. In equal temperament, frequencies of neighboring tones differ by the irrational factor $2^{1/12}$ so that
they are equidistant in a logarithmic representation (see Fig.~\ref{figscale}c). However, this invariance under key shifts comes at the price that all intervals
(except the octave) are slightly out of tune, but apparently our civilization learned to tolerate these discrepancies.

\section{Inharmonicity}

\begin{figure}[t]
\centering\includegraphics[width=150mm]{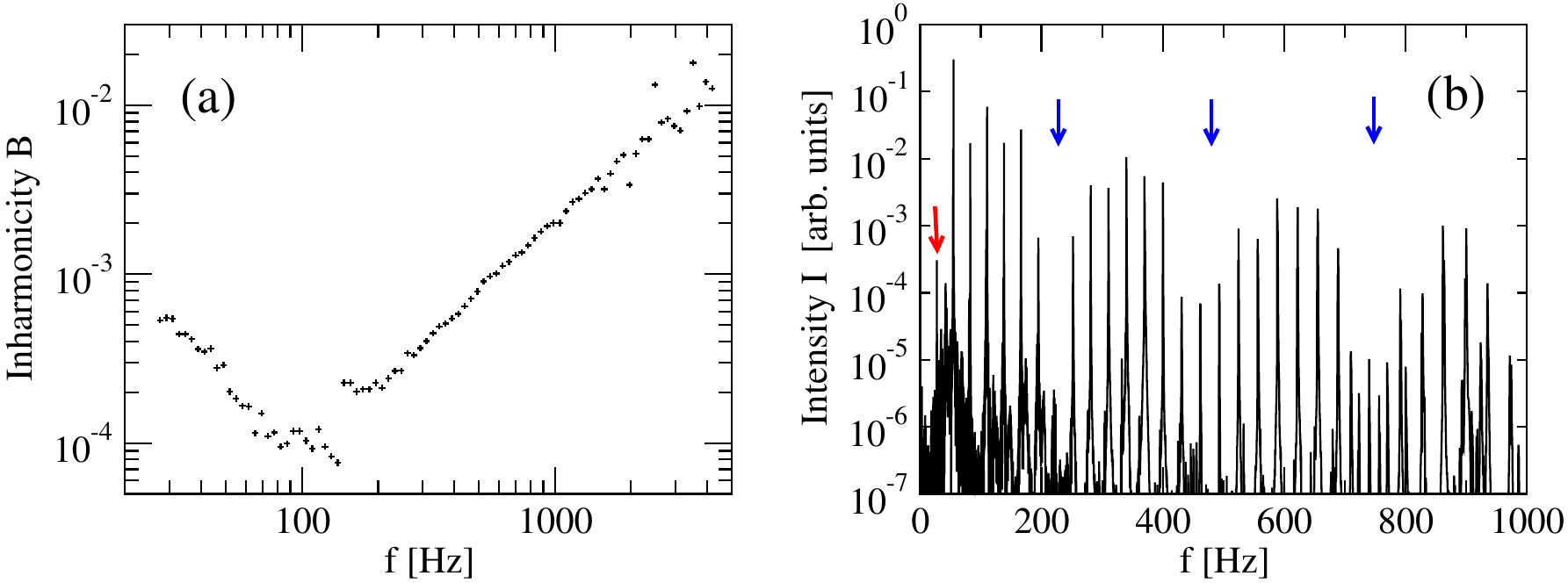}
\caption{\small Left: Inharmonicity coefficients $B$ of an upright piano. The two parts of the data correspond to the two diagonally crossed bass and treble
sections of the strings. Right: Power spectrum of the leftmost string. The red arrow marks the fundamental frequency of 27.5 Hz. The blue arrows indicate particularly
weak partials which are suppressed due to the position of the hammer.}
\label{figpiano}
\end{figure}

The harmonic series of overtones $f_n = n f_1$ is valid only for ideal oscillators whose evolution is governed by a linear second-order partial differential equation. In realistic
musical instruments there are higher-order corrections in the force law which lead to small deviations from the harmonic spectrum. The degree of inharmonicity is
characteristic of each instrument and accounts for much of the color and texture of its sound.

Inharmonicity in string instruments is caused by the circumstance that a realistic string is an intermediate between an ideal string and a stiff bar. An ideal string
vibrates according to the differential equation $\ddot y \propto - y''$ with a linear dispersion $f\propto|k|$, while a stiff bar is known to evolve according to a
fourth-order differential equation $\ddot y\propto -y''''$ with a quadratic dispersion $f \propto k^2$. Therefore, the stiffness of realistic strings causes
lowest-order corrections of the form
\begin{equation}
\ddot y \propto - y'' - \epsilon y'''' \qquad \Rightarrow \qquad f^2 \propto k^2 + \epsilon k^4
\end{equation}
so that the spectrum of a string is given by
\begin{equation}
\label{Inharmonicity}
f_n  \;\propto \; n \, f_1 \,\sqrt{1 + Bn^2 } \,,\qquad n=1,2,\ldots
\end{equation}
where $f_1$ denotes the fundamental frequency and $f_n$ is the frequency of the $n^{\rm th}$ partial. The dimensionless number $B$ is the
so-called \textit{inharmonicity coefficient} which depends on the length, diameter, tension and material properties of the string. In a piano the value
of $B$ varies typically between 0.0002 for bass strings up to 0.4 for treble strings (see left panel in Fig.~\ref{figpiano}). Strong inharmonicity can
be recognized as an unpleasant dulcimer-like sound. An important part in the art of piano construction is to keep the inharmonicity as small as possible.

\section{Perception of being in tune}

\begin{figure}[t]
\centering\includegraphics[width=150mm]{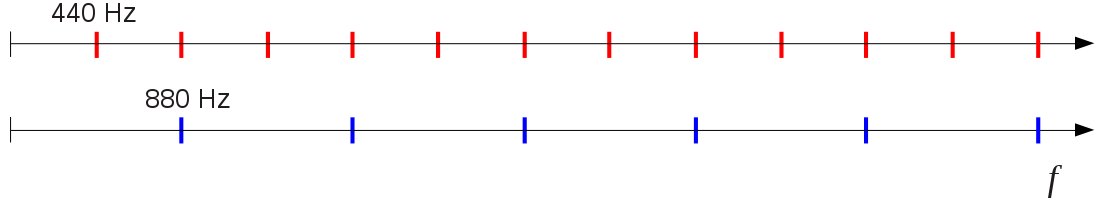}
\caption{\small Harmonic spectrum of partials of an octave in a linear representation. The octave is perceived as pleasant since every second partial of the
lower tone locks in with one of the a partials of the upper one.}
\label{octave}
\end{figure}

As mentioned before, intervals with simple rational frequency ratios are perceived as pleasant. In this context it is important to note that the human ear,
when hearing two different tones simultaneously, cannot evaluate the frequency ratio of the fundamental modes directly: rather it recognizes coincidences
in the corresponding overtone spectra. For example, if we hear an octave, say A2-A3, our ear compares the partials 2,4,6,... of the lower tone with the
partials 1,2,3... of the upper one, perceiving the octave as being `in tune' when the two harmonic series lock in (see Fig.~\ref{octave}). 

\begin{figure}[b]
\centering\includegraphics[width=155mm]{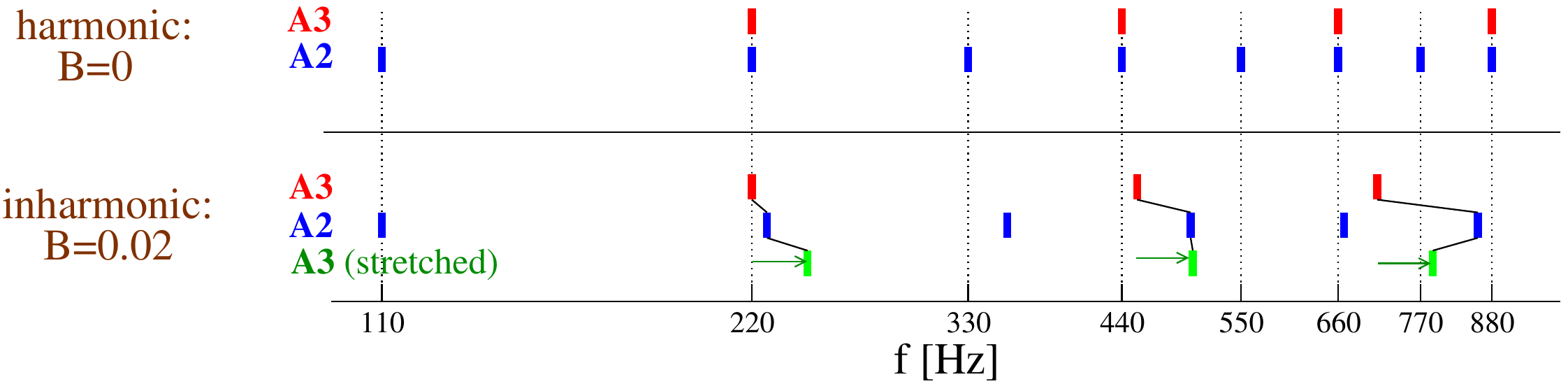}
\caption{\small Compensating inharmonicity by stretching octaves (see text).}
\label{figstretch}
\end{figure}

In the presence of inharmonic corrections, however, it is no longer possible to match the two series exactly. In this case our ears search for the best possible
compromise, minimizing the frequency differences between almost coinciding low-lying partials. These tiny differences are heard as so-called \textit{beats},
a superposition of enveloping modulations of a few Hertz which an aural tuner tries to make as slow as possible. As demonstrated in Fig.~\ref{figstretch},
in an octave such a compromise can be achieved by slightly increasing the pitch of the upper tone. This means that we perceive an interval as correctly tuned
if it is slightly out of tune in the mathematical sense. This correction, called \textit{stretch}, plays a major role in the practice of tuning, even when the
inharmonicity of the instrument is small.

Today high-priced electronic tuning systems are available which can compute the appropriate stretch for individual instruments. To this end some of the tones are
recorded and the inharmonicity coefficients of the strings are estimated by identifying the partials in the corresponding spectra. The stretch is then computed
e.g. by selecting  sequences of octaves and stretching them in such a way that the fourth partial of the lower tone coincides with the second partial of the upper
(4:2-tuning). More specifically, enumerating the piano keys by $k=1\ldots K$ and denoting by $f_n^{(k)}$ the $n^{\rm th}$ partial of the $k^{\rm th}$ string,
this 4:2 method provides $K-12$ equations for the octave stretches of the form
\begin{equation}
\frac{f_1^{(k+12)}}{f_1^{(k)}}\;=\; \frac{r_4^{(k)}}{r_2^{(k+12)}}\,,
\end{equation}
where $r_n^{k}=f_n^{(k)}/f_1^{(k)}$ is the ratio of the $n^{\rm th}$ partial to the fundamental frequency of the string. By taking the logarithm this turns
into a system of $K-12$ linear equations for the $K$ unknown fundamental frequencies $f_1^{(k)}$. The remaining 12 unknowns are determined by the reference
pitch A440 and the choice of the temperament. Equal temperament can be approximated by introducing a quadratic penalty function for the change of adjacent
interval ratios. By solving these equations one can translate the measured partials directly into a tuning curve. If the inharmonicity coefficient is a
piecewise smooth function (like the one shown in Fig.~\ref{figpiano}), the tuning curve will be piecewise smooth as well. Likewise one can use a 6:3 tuning
scheme, which produces an even larger stretch. The overall magnitude of the stretch is therefore not strictly defined but rather a matter of taste. Some
devices even interpolate between 4:2 and 6:3 stretching in order to get a more acceptable compromise.

Computing the stretch by direct comparison of partials as described above yields piecewise smooth tuning curves. However, as mentioned in the Introduction,
aural tuners produce tuning curves with pronounced fluctuations on top of the overall stretch, especially in the bass and in the treble. One of the main messages
of this work is the conjecture that these fluctuations are not random but to some extent essential for a good tuning result. 

The fluctuations may have different reasons. On the one hand, each partial couples differently to the resonator of the instrument (the soundboard of the piano),
leading to additional frequency shifts so that the inharmonic spectrum deviates slightly from the predicted form in Eq.~(\ref{Inharmonicity}). Another reason
is the highly irregular intensity of the partials. As shown in the right panel of Fig.~\ref{figpiano}, the spectrum of a piano string may consist of dozens of
partials, but even adjacent partials may differ in their power by more than a magnitude. Even worse, some of the partials (indicated by blue arrows in the figure)
are strongly suppressed if the hammer hits the string at a node of the corresponding vibrational mode. This suggest that in realistic situations the perception
of being in tune does not only depend on the frequency of the partials but also on their amplitude.

\section{Psychoacoustic aspects}

As tuning can be understood as the search for a compromise in matching higher partials, it will significantly depend on the acoustic and psychoacoustic
properties of the inner ear. Psychoacoustics is a research field on its own (see e.g. \cite{FastlZwicker,PlackEtAl,MooreGlasberg}) and plays an important
role e.g. in lossy data compression methods such as MP3. Here we only sketch a few basic elements which are essential for the method presented below.

Let us first consider the frequency range of the ear. Starting point is a sound wave, which can be described as a time-dependent pressure variation $p(t)$.
Its complex-valued Fourier transform is given by
\begin{equation}
\tilde p(f) = \frac{1}{\sqrt{2\pi}} \int {\rm d}t \, e^{2\pi i f t} \,  p(t)\,,
\end{equation}
where $\tilde p(-f) = \tilde p^*(f)$. The corresponding power spectrum
\begin{equation}
I(f) = |\tilde p(f)^2| 
\end{equation}
describes the energy density of the spectral line at frequency $f$. As a technically useful measure one defines the logarithmic \textit{sound pressure level} (SPL) 
\begin{equation}
\label{form1}
L(f) = 10 \log_{10} \Bigl( \frac{I(f)}{I_0} \Bigr)
\end{equation}
measured in decibels (dB), where $I_0$ refers to the hearing threshold. 

Depending on the frequency the SPL will be correlated with a certain mechanical response in the inner ear. Since the physical transmission mechanism
is highly complex, one usually approximates this relationship by certain weighting functions. Below 55 dB the most commonly used one is the
so-called \textit{A-weighting} according to the international standard IEC 61672:2003 with the filter function
\begin{equation}
\label{form2}
 R_A(f)= {12200^2  f^4\over (f^2+20.6^2)(f^2+12200^2)\,\sqrt{(f^2+107.7^2)\,(f^2+737.9^2)}}
\end{equation}
which defines the A-weighted sound pressure level (SPLA)
\begin{equation}
\label{form3}
L_A(f) = \Bigl(2.0+20\log_{10} R_A(f) \Bigr)\; L(f) 
\end{equation}
in units of A-weighted decibels (dBA). The SPLA can be considered as a rough measure of frequency-dependent energy deposition in the cochlea.

The receptor cells in the inner ear convert the excitation pattern into a certain neuronal response which causes the auditory perception in
the brain. The neuronal processing is even more complex and not yet entirely understood. For this reason one uses a psychoacoustic measure
for the perceived intensity, the so-called loudness $N(f)$, which is an empirical psychological quantity averaged over many test persons.
According to the literature this relationship is well approximated by a piecewise exponential function and a power law:
\begin{equation}
\label{form4}
N(f) =\left\{
\begin{array}{cc}
       2^{(L_A(f)-40)/10} & \mbox{ if } L_A(f) > 40 {\rm dBA} \\
       (L_A(f)/40)^{2.86} & \mbox{ if } L_A(f) \leq 40 {\rm dBA} 
\end{array}
\right.
\end{equation}
Not only the sensitivity of the ear is frequency-dependent but also its ability to discriminate between different frequencies. In the literature
different measures for the frequency resolution are reported, of which the so-called \textit{just noticeable difference (jnd)} plays the role of
a lower bound~\cite{FastlZwicker}. The \textit{jnd} is usually approximated by
\begin{equation}
\label{jnd}
\Delta f \;=\; \left\{
\begin{array}{cc}
3 \ {\rm Hz} & \mbox{ if } f \leq 500\, {\rm Hz} \\
0.006 f    & \mbox{ if }f > 500\, {\rm Hz}\,.
\end{array}
\right.
\end{equation}
%

\section{Entropy-based tuning scheme}

\begin{figure}[t]
\centering\includegraphics[width=155mm]{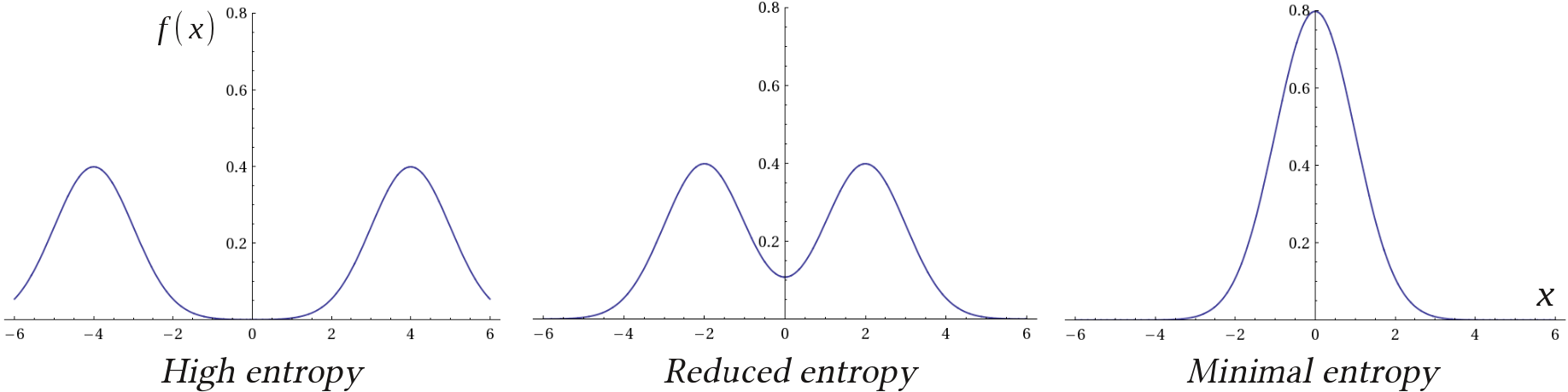}
\caption{\small Shannon entropy as a measure for the coincidence of spectral lines. The figure show the superposition of two Gaussian functions representing two partials. If the two partials are sufficiently separated, the (continuous) entropy $H=-\int_{-\infty}^{+\infty} f(x) \log_2(f(x)) {\rm d}x$ gives a constant value $H\approx 4.094$. When the two partials begin to overlap (audible as \textit{beats}), the entropy decreases and reaches a minimum ($H\approx 2.094$ in this example) if they coincide.}
\label{figentropy}
\end{figure}

We now suggest a simple entropy-based tuning scheme for musical instruments. It is motivated by the observation that tuning can be understood
as the search for the best possible compromise in matching higher partials, and the main idea is that this compromise is characterized by a
local minimum of the entropy of the intensity spectrum. This is highly plausible since the entropy of two spectral lines decreases as they begin to overlap (see Fig.~\ref{figentropy}).

\newpage

\begin{wrapfigure}{r}{90mm}
\vspace{-2mm}
\begin{center}
\includegraphics[width=88mm]{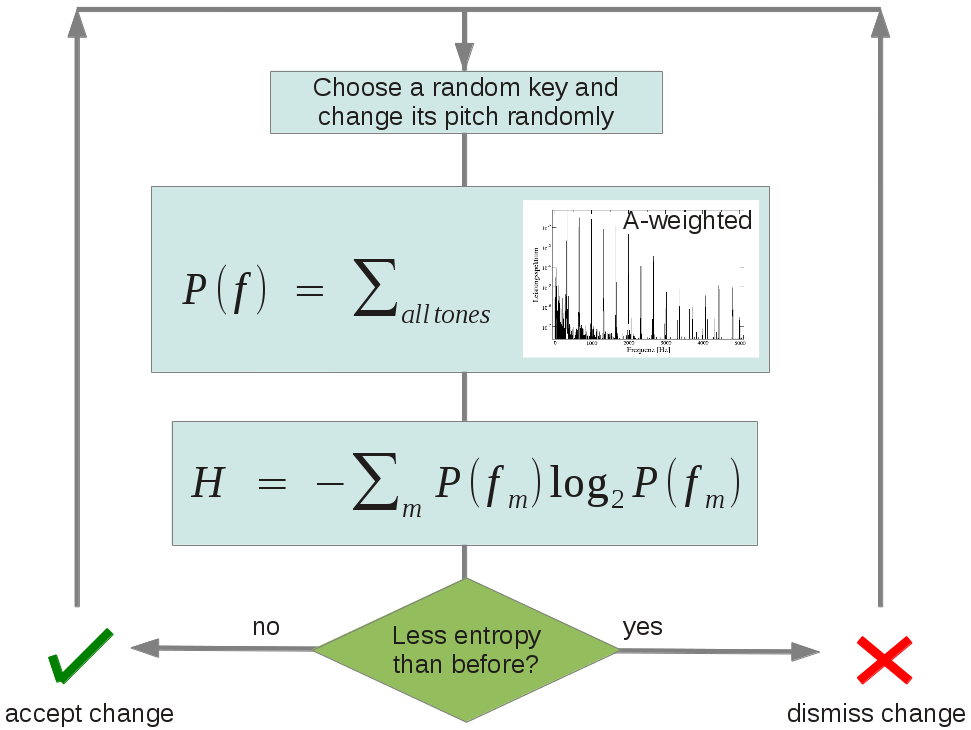} 
\caption{Monte-Carlo scheme.\label{mc}}
\end{center}
\vspace{-13mm}
\end{wrapfigure}

To test this idea we individually recorded all keys of an aurally tuned piano, computed their power spectra and reorganized them in logarithmic bins in order to account
for the finite frequency resolution of the inner ear. Furthermore, we removed the pitch differences, resetting the tune of all tones to equal temperament.
Then we applied the following simple zero-temperature Monte-Carlo scheme of statistical physics (see Fig.~\ref{mc}, technical details are given in the appendix):

\begin{itemize}
\item Add the A-weighted power spectra of all 88 tones and compute the entropy.
\item Randomly change one the of the pitches and compute the entropy again.
\item If the entropy is lower accept the pitch change, otherwise restore the previous value.
\end{itemize}
This simple procedure is iterated until no further improvement is obtained, meaning that the algorithm has found a local minimum of the entropy. Note that by adding up all tones, the method is inherently sensitive to all intervals, not only to octaves.

\section{Discussion}

\begin{figure}[b]
\centering\includegraphics[width=95mm]{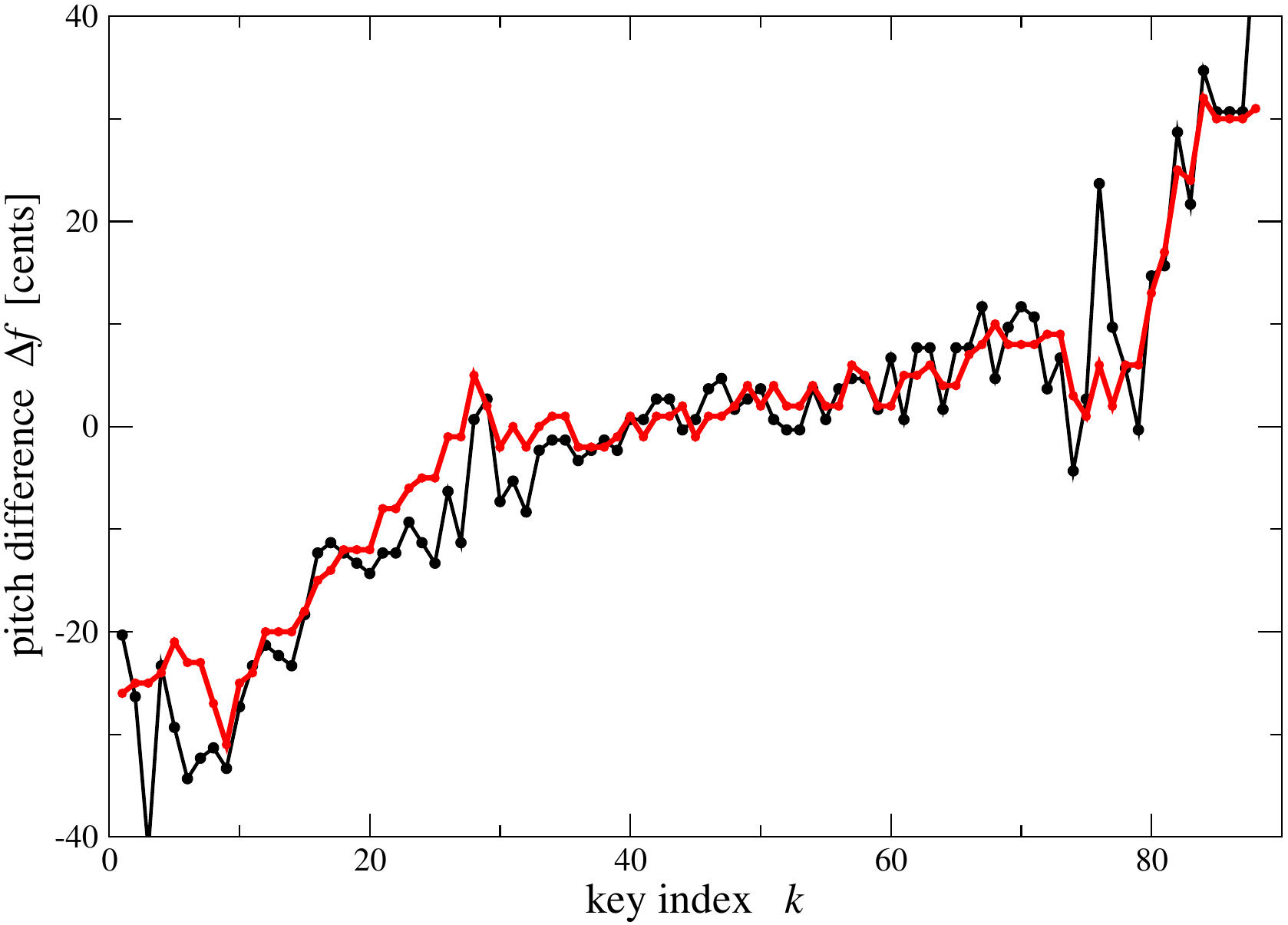}
\caption{\small Typical result of the tuning procedure described in Section 6 (red curve) compared with the tuning curve produced by aural tuning (black curve).}
\label{figresults}
\end{figure}

Fig.~\ref{figresults} shows the resulting tuning curve of a typical run compared with the actual curve produced by an aural tuner for an upright piano. As can
be seen, not only the overall stretch is predicted correctly but even the fluctuations of the two curves are highly correlated, especially in the bass and the 
treble. Apparently the entropy-based tuning method is capable of generating the same individual deviations from the average stretch as an aural tuner. This
is surprising and not yet understood, but it indicates that these fluctuations are reproducible and may play an essential role in the practice of tuning.

The implementation of the method is very easy. The tones are recorded, Fourier-transformed, mapped pointwise by the psychoacoustic filtering functions as described
above, binned logarithmically, added up, and finally plugged into the entropy functional. An explicit identification of higher partials and the measurement of the
inharmonicity is not needed. The method is expected to take automatically any anomalous spectral properties of the instrument into account.

However, the method suggested here is still in an immature state. It could be modified in various respects and a systematic study is still outstanding.
Moreover, the method was tested so far with only one instrument. The main open questions are the following:

\begin{itemize}
\item Apparently there are many local minima, so that the algorithm outlined above produces similar but not reproducible results.
\item The Monte-Carlo results presented above were based on the A-weighted spectra (SPLA) in Eq.~(\ref{form3}). If one uses instead the loudness
defined in Eq.~(\ref{form4}), one obtains unreasonably stretched tuning curves in the bass. 
\item The spectra were logarithmically binned in units of one cent. This models a frequency resolution of one cent, which is smaller than the just notable
difference (\textit{jnd}). However, convolving the spectra with a frequency-dependent Gaussian according to the expected \textit{jnd} in Eq.~(\ref{jnd})
does not improve the results.
\item More advanced Monte Carlo techniques such as simulated annealing have not yet been tested.
\item Instead of adding up the spectra of \textit{all} piano keys, we tried to work with subsets of octaves, fifths and fourths, imitating the practice of aural tuners.
This destabilizes the method, probably driving the pitches out of equal-tempered into just intonation. Apparently the summation over all keys allows the
system as a whole to stay in equal temperament.
\end{itemize}

\noindent
Regarding possible technological realizations, it could be interesting to develop electronic tuning devices using a hybrid method, which first compute an
approximate tuning curve by matching higher partials, and then optimize the fluctuations of the pitches by searching for a suitable local minimum in the vicinity. \\

\noindent\textbf{Acknowledgments}\\
The author thanks for the warm hospitality at the Universidade Federal do Rio Grande do Sul (UFRGS) in Porto Alegre, Brazil, where parts of this work have
been done. This work was supported financially by the German Academic Exchange Service (DAAD) under the Joint Brazil-Germany Cooperation Program PROBRAL.

\appendix
\section{Technical details}

\paragraph{Data recording and preprocessing}

\begin{enumerate}

\item 
Record the piano keys $k=1 \ldots K$ in WAV format. Extract the binary PCM amplitudes and convert them to a series of floating point numbers
$y^{(k)}_j \in \mathbb{R}$ with an index $j=0\ldots ST-1$, where $S=44100 $ Hz is the sample rate and $T\approx 20 s$ is the recording time.
\item 
Apply a fast Fourier transform (e.g. package {\tt fftw3}) to obtain the spectra $\tilde y^{(k)}_q \in \mathbb{C}$ indexed by $q=0\ldots Q$,
where $Q=ST/2$ (the other half of the data is complex conjugate). The $q^{\rm th}$ component corresponds to the frequency $f(q) = q / T$. 
\item
For each $k$ coarse-grain the power spectrum $|\tilde y^{(k)}_q|^2 \in \mathbb{R}^+$ by logarithmic binning. To this end define an
array $I^{(k)}_m\in\mathbb{R}^+$ corresponding to the frequencies $f(m)=10 \cdot 2^{m/1200}$ Hz with $m$ running from zero (10 Hz) to 12000 (10 kHz). Let
  \begin{equation}
  I^{(k)}_m := \sum_{q=0}^Q \delta_{m, [1200+\log_2(\frac{q}{10T})]}\,  |\tilde y^{(k)}_q|^2 \,,
  \end{equation}
where $[\cdot]$ denotes rounding to an integer. Note that in this representation adjacent bins differ by a frequency ratio of one cent.
\item Map the intensities $I_m^{(k)}$ to the corresponding A-weighted sound pressure levels (SPLA) $L_m^{(k)}$. 
\item
After this preprocessing the pitch of a tone can be increased or lowered by $c$ cents through a shift of the array index $m \to m-c$. This allows us to tune the instrument virtually on the computer. Remove the recorded pitch deviations by tuning the first partial to equal temperament, i.e. $f_1^{(k)}=440\cdot 2^{(k-k_0)/12}$ Hz, where $k_0$ is the index of A440. This means that the recorded stretch corrections are initially eliminated.
\end{enumerate}

\paragraph{Monte-Carlo dynamics}
\begin{enumerate}
\item Change one of the pitch differences randomly by $\pm 1$ cent.
\item Compute the sum $p_m = \sum_{k=1}^{88} L_m^{(k)}$ of the SPLA over all keys.
\item Normalize $p_m$ such that $\sum_m p_m=1$.
\item Compute the Shannon entropy $H=-\sum_m p_m \ln p_m$.
\item If the entropy decreases, keep the pitch change, otherwise restore the old pitch.
\end{enumerate}

\noindent
This procedure is repeated until no further changes take place.\\

\section*{References}


\begin{thebibliography}{99}
\bibitem{roederer} J. G. Roederer, \textit{Introduction to the Physics and Psychophysics of Music}, Springer, New York (1973).
\bibitem{railsback} O. L. Railsback, \textit{Scale Temperament as Applied to Piano Tuning}. Journal of the Acoustical Society of America 9 (3): 274 (1938).
\bibitem{figuresource}
Figure taken from: http://en.wikipedia.org/wiki/File:Railsback2.png (March 2012).
\bibitem{figuresource2}
Figure taken from: http://en.wikipedia.org/wiki/File:Harmonic-partials-on-strings (March 2012).
\bibitem{FletcherRossing}
N. H. Fletcher and T. D. Rossing, \textit{The Physics of Musical Instruments}, Springer, New York (1991).
\bibitem{FastlZwicker} H. Fastl and E. Zwicker, \textit{Psychoacoustics: facts and models}, Springer, New York (2007).
\bibitem{PlackEtAl} C. J. Plack, A. J. Oxenham, and  R. F. Richard, eds.  \textit{Pitch: Neural Coding and Perception}. Springer, New York (2005).
\bibitem{MooreGlasberg} B. C. Moore and B. R. Glasberg,  \textit{Thresholds for hearing mistuned partials as separate tones in harmonic complexes}. J. Acoust. Soc. Am., 80, 479–483 (1986).
\end{thebibliography}
\end{document}